\documentclass[a4paper]{jpconf}
\usepackage{graphicx}
\usepackage{amsmath}

\usepackage{subfig}
\usepackage{hyperref}
\hypersetup{colorlinks=true,allcolors=black}
\usepackage[nolist]{acronym}
\usepackage[font=footnotesize,labelfont=bf]{caption}
\usepackage{multirow}

\usepackage{fancyhdr}

\usepackage{tikz}
\usetikzlibrary{spy}
\begin{document}

\title{Evaluating Global Blockage engineering parametrizations with LES}

\author{G Centurelli\textsuperscript{1}, L Vollmer\textsuperscript{2} , J Schmidt\textsuperscript{2}, M D\"orenk\"amper\textsuperscript{2}, M Schr\"oder\textsuperscript{1,2}, L J Lukassen\textsuperscript{1} and J Peinke\textsuperscript{1}}
\address{\textsuperscript{1} ForWind, Carl von Ossietzky University Oldenburg, Küpkersweg 70, 26129 Oldenburg, Germany}
\address{\textsuperscript{2} Fraunhofer IWES, K\"upkersweg 70, 26129 Oldenburg, Germany}

\ead{gabriele.centurelli@uni-oldenburg.de}

\begin{abstract}
Under the term \textit{global blockage}, the cumulative induction of wind turbines in a wind farm has been recently suspected to be responsible for observed overestimations of the energy yield in large-size wind farms.
In this paper, the practice of modeling this effect after linear superposition of single turbine inductions, calculated with three of the most recent analytic models, is compared to Large-Eddy-Simulations of wind farms. 
We compare the models across two different farms, composed of 9 and 49 turbines, with two different heights of the atmospheric boundary layer, 300 and 500\,m. The results show that the differences between the analytical models are negligible while they substantially differ from the LES results. The linear superposition of induction consistently underestimates the velocity deficit in front of the farm with an error that increases as the wind farm size grows and the ABL height decreases. Also, when calculating the power output at the turbines of the farm, all the analytical models considered do not agree with the LES. 
These comparisons reveal that the farm interactions with the atmospheric boundary layer may highly outclass the turbine induction in determining the extent of the global blockage effect. Therefore, we present a first dimensional approach to the problem based on LES, aimed at simplifying its characterization.

\end{abstract}

\section{Introduction}

Despite being already conceived in the very early theory of Betz, it is only in recent years that the induction of a wind turbine was finally given a representation 
in engineering models that aim at modelling the flow inside of wind farms. 
There is, in fact,
a growing awareness that upstream interactions between wind and wind turbines should no longer be neglected \cite{Bleeg2018}.  Supporting this theory, there is consistent evidence of wind speed deceleration upstream of wind farms suspected to cause efficiency losses, especially of the first turbines row \cite{orsted2019}. This effect, discussed under the name of wind farm blockage or global blockage, is at our day and age still debated in science and industrial applications, and different postulates on its origin exist. \\
An early study on wind farm power losses caused by inflow wind speed reduction can be found in \cite{Smith2010}, in there the wind farm is suspected to generate atmospheric gravity waves, similarly to mountains, capable of reducing the inflow speed through pressure perturbation. 
The same idea has been more recently extensively investigated by Allaerts and Meyers in \cite{Allaerts2017,Allaerts2018,Allaerts2019}, where by means of Large-Eddy Simulations (LES) the authors were able to better describe the extent of the phenomena and to connect it to the stratification of the \ac{ABL}. 
A more general discussion on the existence of global blockage and its effect on power production can be found in \cite{Bleeg2018}. In this work, the authors compared the wind speed measurements collected by different met-masts before and after a nearby wind farm was installed. The wind speed was observed mostly to decrease in the second period even when the met-masts measured 2$\div$ 10 rotor diameters upstream the wind farm.
Other, more specific, \emph{in situ} evidence of the phenomena can be found in the measurements with remote sensing technique of Asimakopoulos et al. \cite{Asimakopoulos2014} and especially Schneemann et al. \cite{Schneemann2021}. In this novel work, the wind speed at transition piece height ($\approx$25\,m above mean sea level) is found to decrease up to 4.5\,\% between 40 and 10 rotor diameters upstream of the \emph{Global Tech 1} wind farm when stable stratification characterizes the \ac{ABL}. 
\\
As the pool of evidence for global blockage is growing larger, the discussion on whether and how to consider its effect in wind farm energy yield assessment is still quite open. 
\\
An important contribution in this direction is given by the work of Segalini and Dahlberg \cite{Segalini2020}. They derived an empirical correlation describing velocity deficit for the first row at varying farm layout, from a series of wind tunnel measurements. A different approach can be found in \cite{Branlard2020,Nygaard2020}. In both publications, the authors propose to model the global blockage effect by linear superposition of the induction effect of the single turbines, in a similar fashion to what is currently done for wake modeling.

Nygaard et al. \cite{Nygaard2020} also attempted a validation of the method. In there, the power output of the first row at the UK-based \emph{Gunfleet Sands} wind farm is simulated for two particular wind direction bins at a fixed velocity. By comparing the results with averaged \ac{SCADA} data, the authors observe that ``in its present form the model is likely missing some element of the relevant physics''. This important conclusion cannot, unfortunately, be considered definitive. The wind farm chosen is surrounded either by the coast or by nearby farms, in this condition the blockage effect is not the only factor determining wind velocity distributions, and phenomena such as external wind farm wakes or coastal wind speed gradients may even play a greater role.
Within this work, we want to contribute in assessing how much the linear superposition of single turbine induction should be expected to suffice in the modeling of wind speed deceleration and hence wind farm power reduction connected to the effect of global blockage. As reference for the comparison we use LES capable of describing the development of the atmospheric boundary layer under physical sound conditions. With the numerical environment we remove parasite perturbations 
that may compromise the fairness of the comparison in measurement-based validation approaches.\\
After a brief description of the induction models, their implementation in the Fraunhofer IWES in-house software \emph{flappy} (Farm Layout Program in Python), and the LES  set-up in section \ref{s:2}, we present results for the comparison models/LES on wind velocity and power production in section \ref{s:results}. 
As a closing argument, in section \ref{s:dimension}, we propose a simple dimensional study to characterize the wind velocity deficit the LES predict in front of the farm, seeing the necessity of improving the current models.

\section{Models Description and set-up}\label{s:2}

Within the last years a good number of engineering models describing the velocity deficit in the induction zone of a wind turbine have been proposed. Branlard et al. \cite{BranMay2020} review most of them, offering a comparison of their performance against an \ac{AD} \ac{RANS} simulation. The \ac{VC} model of \cite{Branlard2020} appears as the one offering the best accuracy.
In this manuscript, we consider the \ac{VC} model and two additional models with lower computational effort: the \emph{Self-Similar} model of Troldborg et al. \cite{Troldborg2017} and the potential flow model of Gribben and Hawkes \cite{Gribben2019}, here as \emph{Rankine Half-Body}.

\subsection{Vortex Model (VC)}
Inspired by the cylinder model for impellers of Joukowski, the \textit{Vortex Cylinder} model \cite{Branlard2020} computes the perturbation velocity components by linear superposition of the effects from the three main vortexes generated in the interaction between wind and turbine blades: the straight root vortex, the straight bound lifting lines at the blades, and helical vortex trailed at the blades tip. As an infinite number of blades is assumed, the lifting lines at the rotor disk merge into a bound vortex sheet and the helical vortex can be decomposed in line vortex filaments and a circular component responsible for a tangential circulation, $\gamma_t$ \cite{Branlard2014}.\\
The computation of velocity perturbation happens thanks to the Biot-Savart law integration applied to any of the vortex systems defined. Multiple concentric cylinders supporting $\gamma_t$ may be considered to describe the wake expansion, but, as done in \cite{Branlard2020,BranMay2020}, only the single cylinder matching the rotor dimension is considered here.\\
Under the previous assumptions, only the tangential circulation is the capable of providing the axial induction. Therefore it is possible to connect $\gamma_t$ to the induction factor, $\bar{a}$, thanks to the relation $\gamma_t = -2\, U_{ref}\,\bar{a} $, where $U_{ref}$ is the velocity of the free stream approaching the turbine, as demonstrated in \cite{Branlard2014}. 
In the present work, $\bar{a}$ for the model is computed according to the polynomial expansion on the global thrust coefficient $C_t$,
\begin{equation}
\bar{a} = 0.169C_t + 0.400C^2_t - 0.482C^3_t + 0.396C^4_t
\end{equation}
, found in \cite{Branlard2020} to provide a better agreement with an \ac{AD} RANS throughout the inflow region $x\in [-2.5\,$D,$-0.5\,$D], with D the rotor diameter, with respect to the one-dimension momentum theory.
The \ac{VC} model has been introduced in our wind farm modeling tool implementing the approach followed for FLORIS \cite{FLORIS2020} available at \url{https://github.com/ebranlard/wiz}.

\subsection{Self-Similar model (SS)}
The axial velocity deficit at hub height, $u_x(x,r=0)$, in the \ac{VC} model takes a very simplified formulation that reads as,
\begin{equation}\label{eq:ax_vc_udef}
    u_x(x,r=0) \equiv u_{x,0}=\bar{a}U_{ref}\left( \frac{x}{\sqrt{R^2 + x^2}}\right)
\end{equation}
with $x$ and $r$ being the coordinates of a cylindrical frame or reference centered at the rotor hub height ($z_{HH}$).
Troldborg et al., \cite{Troldborg2017}, have observed the induction profiles for different turbines simulated with RANS to exhibit \textit{self-similar} behaviour when $x > R$. They hence proposed to compute the velocity deficit using (\ref{eq:ax_vc_udef}) and describing the radial decay as
\begin{align*}
\centering 
f(\epsilon) = \frac{1}{cosh^\alpha(\beta\epsilon)}, \quad\quad
\epsilon = \frac{r}{r_{1/2}(x)}, \quad\quad
r_{1/2}(x) =  \sqrt{\lambda\left[\eta - \left(\frac{x}{R}^2\right)\right]}
\end{align*}
with $\alpha=8/9,\,\beta=\sqrt{2},\,\lambda=0.587,\,\eta=1.32$. Differently to what chosen for the \ac{VC} model, the induction factor $\bar{a}$ is based here on the 1\,D momentum theory, $\bar{a} = 0.5\,(1-\sqrt{1-\nu C_t})$ with the constant $\nu=1.1$ providing better agreement with RANS simulations.

\subsection{Rankine Half-Body (RHB)}
Gribben and Hawks in \cite{Gribben2019} suggest that the flow around a wind turbine resembles the response of a uniform flow field to a constant momentum source. In this typical problem of potential flow theory, the main flow is inviscidly deflected around a \textit{Rankine Half-Body}, the imaginary surface described by the main stagnation streamline, described in a spherical frame of coordinates ($r,\theta$) centered at turbine hub height as
\begin{equation}
cos(\theta) - \frac{2\pi U_{ref}}{\sigma}\,r^2\,sin^2(\theta) = -1
\end{equation}
with $\sigma=2\,\bar{a}\,A_{rotor}\,U_{ref}$ being the momentum source, and $\bar{a}$ again derived from the one-dimensional momentum theory.
Any of the cartesian velocity perturbation components is given by
\begin{equation}
\centering
u_k = \frac{\sigma}{4\pi}\frac{k}{\left(\sum_k{k^2}\right)^{3/2}} \quad\quad\quad    k = x,y,z.
\end{equation}

\subsection{Empirical wind speed reduction by Segalini and Dahlberg (Seg)}
As additional model for comparison, we consider the empirical wind speed reduction derived by Segalini and Dahlberg \cite{Segalini2020} (\emph{Seg})
\begin{equation}\label{eq:Seg}
u_{x_0} = U_{ref}\left\{0.097 \left(\frac{\Delta_x\Delta_y}{D^2} \right)^{-0.9} [1 - \exp(0.88 - 0.88\,N_{rows})]\right\}.
\end{equation}
Here, $u_{x_0}$ is the wind speed deficit at the front-most row middle turbine, $\Delta_x$ and $\Delta_y$ the streamwise and spanwise distances between the turbines and $N_{rows}$ the number of streamwise rows of the wind farm. While not explicitly stated in \cite{Segalini2020}, we use here $u_{x_0}$ as initialization velocity at any turbine of the farm.

\subsection{Case study setup and LES}

The framework of our comparison is to perform simulations for the very same wind farm layout either in \ac{AD} LES and in the Farm Layout Program in Python\footnote{\emph{flappy} v0.4.3} (\emph{flappy}) developed  at Fraunhofer IWES \cite{Schmidt2021inpreparation}, the successor of \emph{flapFOAM} (Farm  Layout  Program  coupled  to  OpenFOAM) \cite{Schmidt2014}.
\\
The wind turbines considered in the simulations have a constant $C_t$ of 0.86 and a constant power coefficient. 
Thus, they are characterized by a rather high, but not unusual, induction factor.
The turbine's geometry is typical for offshore application with a rotor diameter (D) of 160\,m, resulting in a rated power of approximately 8\,MW, and a hub height of 110\,m.
Two basic wind farm setups were simulated, one with three turbines in streamwise and spanwise direction, respectively (3\,$\times$\,3), and the other with seven turbines in each direction (7\,$\times$\,7).
The distance between the turbines in the streamwise and in spanwise directions is 7\,D and 5\,D, respectively. Additional simulations with a single turbine and just the first turbine row were conducted to serve as references.\\
The LES data was created with the \textit{Parallelized
Large Eddy Simulations Model} (PALM) \cite{Maronga2020}. 
Further details about the used model setup can be found in \cite{Vollmer2021inpreparation}.
The wind profile and turbulence of the simulations are developed during a model spin-up period of 35 hours.
The stratification of the simulated boundary layer at the start of the spin-up simulation is neutral up to a capping inversion. For simplification, we set the inversion layer to have a strong potential temperature gradient of 8\,K/100\,m and a thickness of 100\,m, above this region a stable stratification of 1\,K/100\,m finds place up to the upper boundary of the domain. Thus, any flow above the inversion layer is basically decoupled from the boundary layer flow. 
\\
Two different wind fields are created for the wind farm simulations in the LES, differentiating in the height of the inversion layer, with the inversion in the ``ABL 300''-simulation starting at 300\,m height and the ``ABL 500''-simulation at 500\,m, thus resulting in four sets of LES reference wind farm simulations: (3$\times$3\_500, 3$\times$3\_300, 7$\times$7\_500 and 7$\times$7\_300). The horizontal domain sizes of the simulations are adapted to the size of the wind farm. The domains extend over approximately 13 $L_x$ in streamwise and 13 $L_y$ in spanwise and over 3500\,m in vertical direction, with $L_x$ and $L_y$ the length and width of the wind farm, respectively.
The average wind speed profiles and hub height turbulence intensity at the inflow boundary serve as input to the engineering model simulations with \emph{flappy}. 
\\
In \emph{flappy} the final flow field is derived in an iterative approach.
Any velocity perturbation computed by induction or wake models are linearly superimposed.  No induction model is allowed to provide a contribution in the projected cylinder at the back of the rotor at which the induction is computed. 
The presence of the ground is mimicked by also modelling a second turbine mirrored at the ground surface. Wake effects are modelled with a modified version of the Bastankhah wake model as described in \cite{Niayifar2015}. However, as the wake modelling only has a marginal influence on the modification of the flow field by the induction models, we will not explore the representation of the wakes further.

\section{Results}\label{s:results}

\begin{figure}[t!]
\centering
\includegraphics[width = \textwidth]{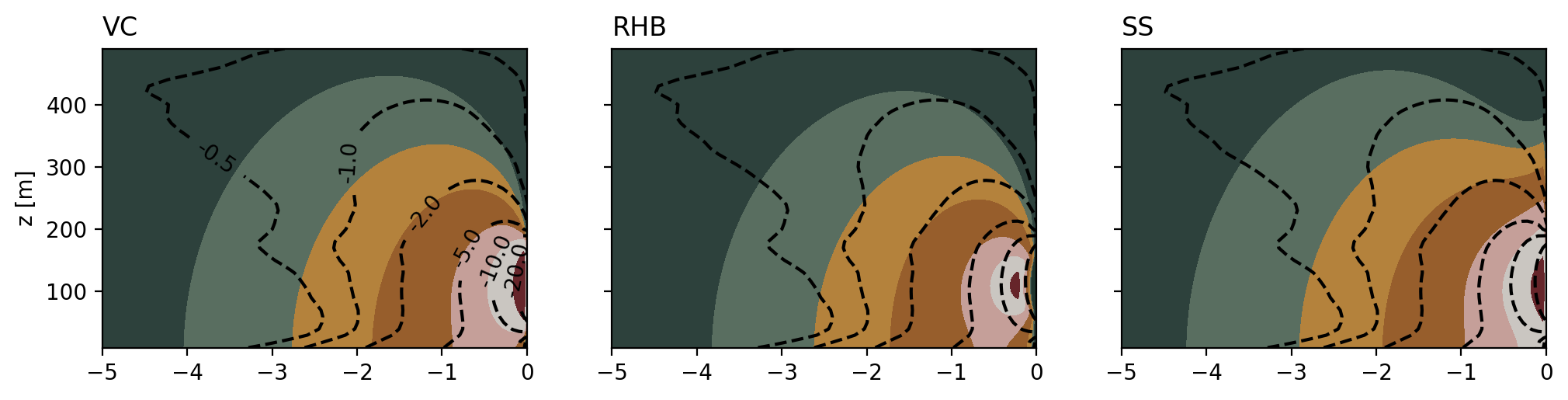}
\includegraphics[width = \textwidth]{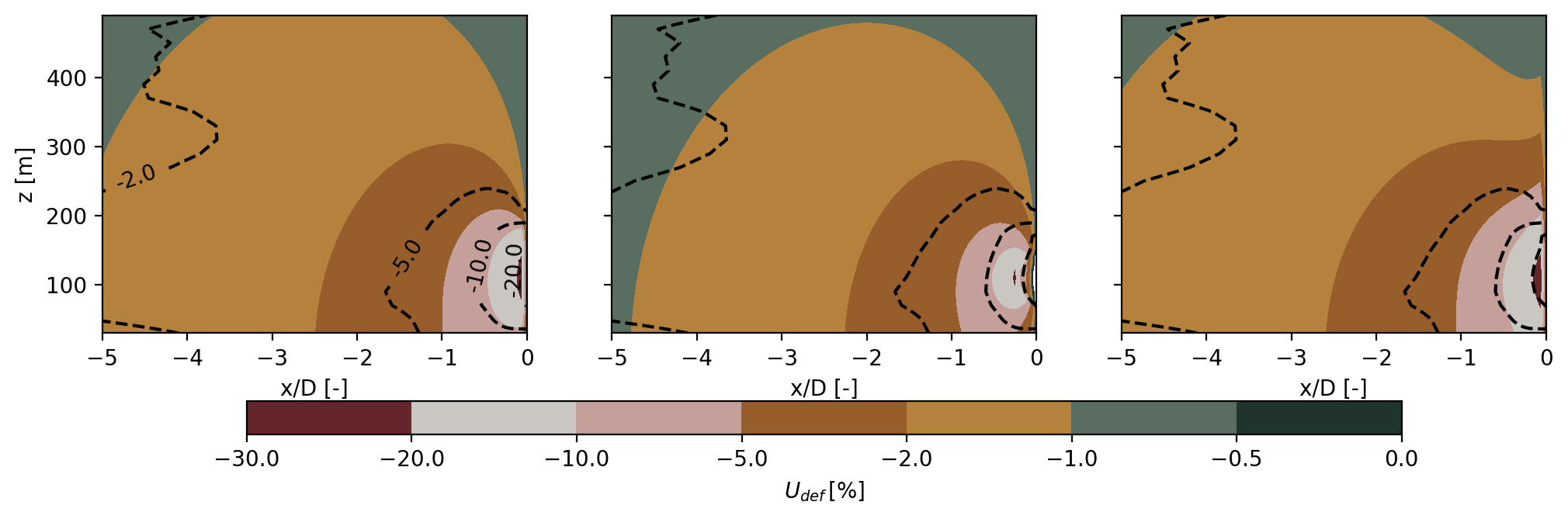}
\caption{Contours of upstream mean velocity deficit averaged in the span-wise direction for the rotor frontal cross section. LES values are the dashed lines. Single turbine with \ac{ABL} 500 (top), middle turbine of 7$\times$7 farm first row with \ac{ABL} 500 (bottom)}
\label{fig:wsz_1}
\end{figure}

In this section, the predictions of the engineering models are compared against the LES results. First we focus mostly on the axial velocity deficit that we define as 
\begin{equation}
    U_{def} = (U_x(x,z) - U_{x,ref}(z)) / U_{x,ref}(z = z_{HH}) \times 100
\end{equation}
, the reference velocity is taken as the spanwise averaged value from the LES at the inlet of the domain. Later, we look at the models effect when applied to wind farm power prediction. In this section (\ref{sec:wf_pow}), the empirical correlation of Segalini and Dahlberg \cite{Segalini2020} will be discussed as well.\\
It is important to mention that the induction models have no dependence on ABL height. Thus, the comparison with the LES of different ABL height serves as an illustration of an effect that is not reproducible with the implemented models.

\subsection{Velocity field prediction}\label{ss:resvp}
Figure \ref{fig:wsz_1} presents the distribution of the velocity deficit in the vertical plane upstream of a wind turbine. We observe a good agreement between models and LES when the single turbine is simulated in isolation, Fig. \ref{fig:wsz_1}(top), confirming the results of Branlard et al. \cite{BranMay2020} in their comparison with AD-RANS simulations. Qualitatively, the trend of the velocity deficit match up quite well in the region $x > -2\,D$, especially for the \ac{VC} model. For further upstream distances the models tend to overestimate the deficit slightly.

However, repeating the same comparison for the middle turbine of the front row of the 7$\times$7\_500 case, Fig. \ref{fig:wsz_1}(bottom), the previous difference are exacerbated, with significant underestimations of the wind speed reduction for all three models. 
While the wind farm in the LES induces wind speed reductions of up to 2\,\% at distances of 5\,D upstream, the reductions in the models do not exceed 1\,\%. Interestingly, a good agreement remains for the near-rotor region in the VC model simulation. Hence we deduce that the linear superposition method fails to describe the cumulative effect of the wind turbines on the flow.

\begin{figure}[t]
\centering
\begin{minipage}{.33\textwidth}
  \centering
  \includegraphics[width=1\linewidth]{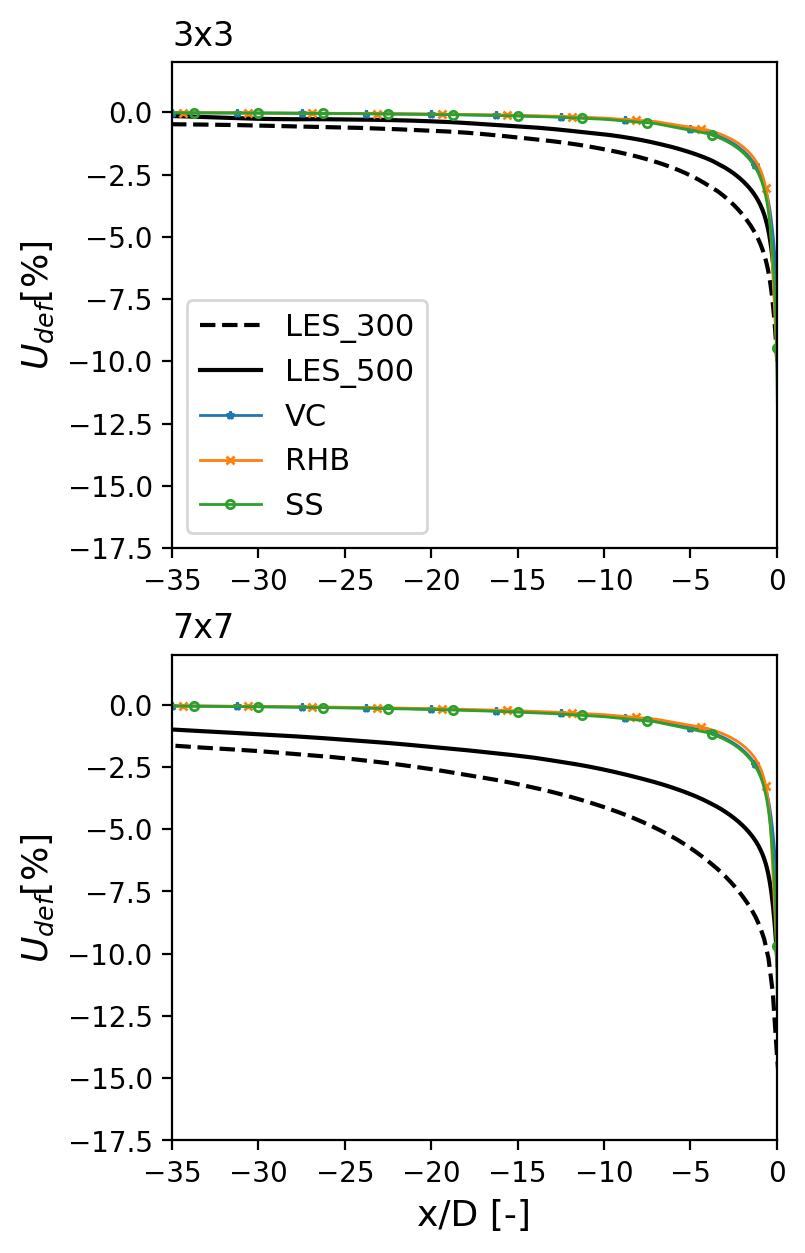}
  \captionof{figure}{Mean Axial velocity components at hub height upstream the wind farms. The value is averaged on the span-wise direction on a window centered on the farm  and $5\,D$ larger of its cross section.}
  \label{fig:ws_x_avg}
\end{minipage}%
\hfill{}
\begin{minipage}{.63\textwidth}
  \centering
  \includegraphics[width=1\linewidth]{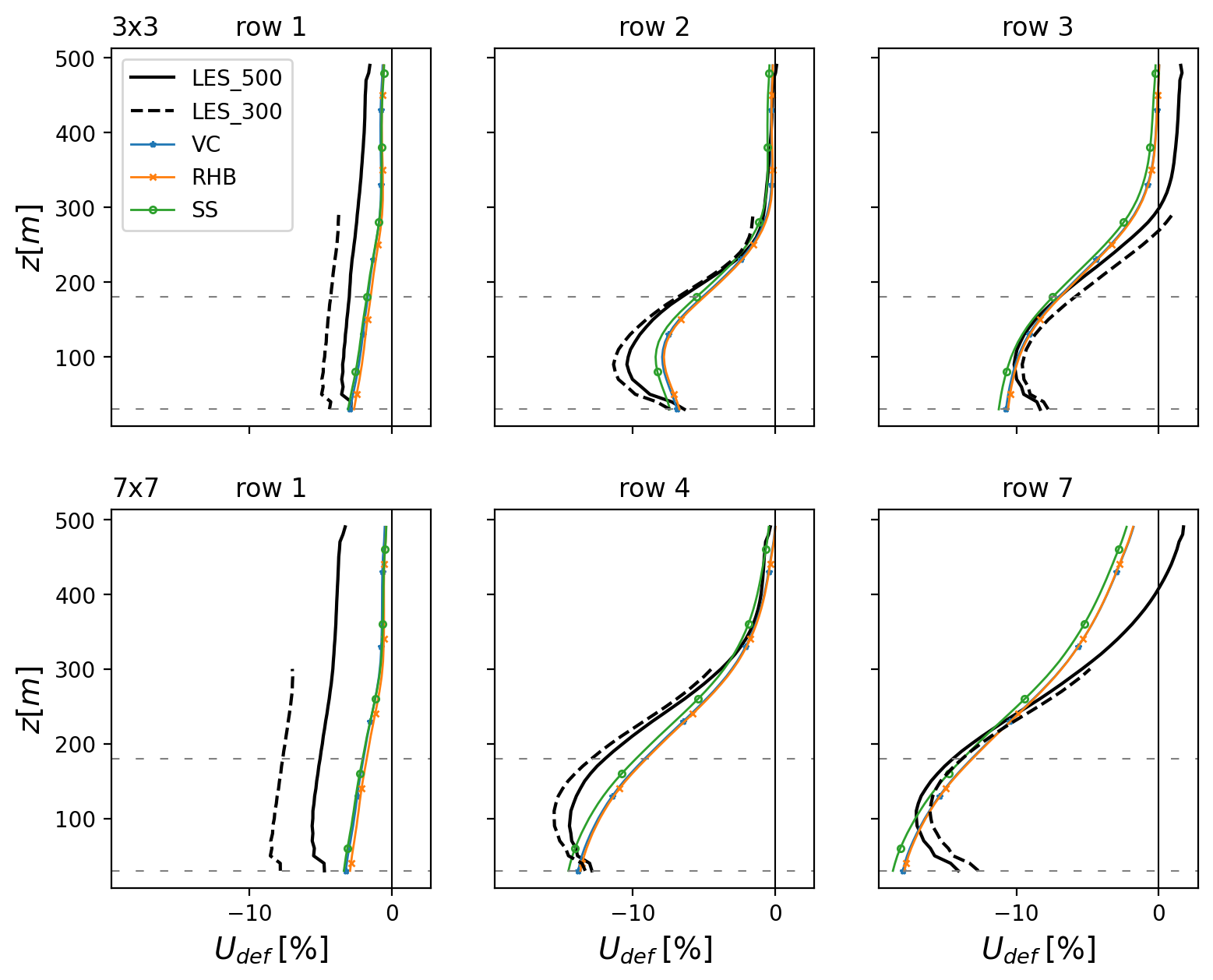}
  \captionof{figure}{Mean axial velocity deficit averaged spanwise at cross planes 1 rotor diameter upstream the first, middle and last row in the farm. Horizontal dashed lines represent the rotor edges.}
 
  \label{fig:ws_z_avg}
\end{minipage}
\end{figure}

Figure \ref{fig:ws_x_avg} provides a different visualisation of the upstream velocity reduction with a display of the mean axial wind speed at hub height.
By comparison to the LES, the models clearly underestimate the velocity deficit throughout the considered domain and compared to the LES results the model difference become marginal. 
From the 3$\times$3 to the 7$\times$7 wind farm, the velocity deficit at $x = -10\,D$ grows of 3 times larger in the LES simulations, while the models predict only a 70\,\% increase.
Similarly, varying the height of the boundary layer strongly affects the inflow velocity deficit at both the farms in the LES. The complete lack of sensitivity on this parameter proves to be critical for the model. 

Given the general good agreement observed for the single turbine case, the results obtained until here point out that there are significant overhead effects contributing to wind speed decelerations in front of the farm that turbine induction cannot justify. 

It is by looking at the vertical distribution of the velocity deficit, Figure \ref{fig:ws_z_avg}, that it becomes even more evident that the inductions effects have too local traits and they cannot fully motivate the global blockage effect. The LES shows the velocity perturbation upstream of the first row to not only be more intense with respect to the models, but also to interest the whole vertical dimension of the ABL.
\begin{figure}[t]
\centering
\includegraphics[width=\textwidth]{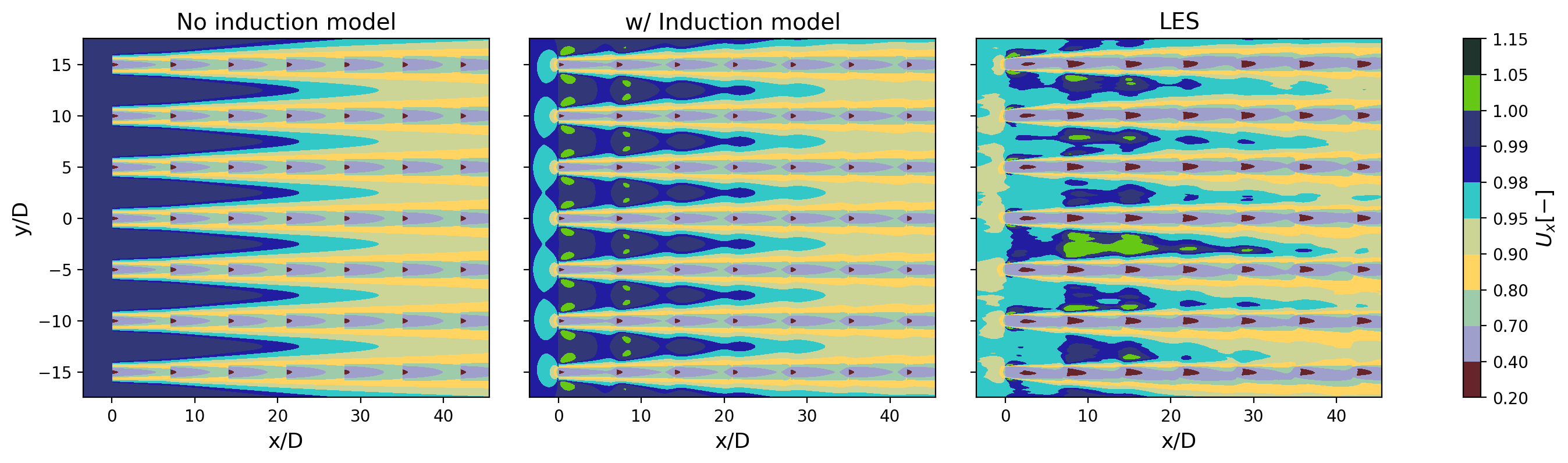}
\caption{Contours of normalized wind speed at hub height for the 7$\times$7\_500 farm. The induction model selected for comparison is the \ac{VC} model.}
\label{fig:ws_conf}
\end{figure}
Figure \ref{fig:ws_z_avg} also displays properties of the flow within the wind farm. 
The vertical profiles of the velocity deficit at the middle and last row in the LES reveal that the initial differences between the two ABL simulated are recovered within the wind farm, and above the rotors the wind speed may even exceed the velocity of the free flow far upstream. This effect is completely outside the analytical models' capability, even the one describing the speed-ups around the rotors.\\\\
To our understanding, the flow patterns observed up to now can only be motivated by pressure perturbations in the lower boundary layer originating, as explained by Allaerts and Meyers in \cite{Allaerts2017}, as a consequence of a displacement of air masses with different density aloft. Clearly, such an effect cannot be replicated with the engineering model approaches where, to derive the perturbation of the velocity field, pressure is considered to be in equilibrium.

As a conclusion to our analysis of the velocity field representation, Fig. \ref{fig:ws_conf} visualises  the horizontal velocity contours at hub height for the 7$\times$7\_500 case. We observe that the use of an induction model enhances the similarity with the results of the LES, especially if the model accounts for lateral speed-ups like the \ac{VC} model used in the comparison. This underlines once more that the models are capable of reproducing the local turbine blockage, but their linear superposition cannot adequately represent the global blockage effect, as the wind farm interactions with the atmospheric boundary layer makes the phenomenon strongly non-linear.

\begin{figure}[t]
\centering
\includegraphics[width=0.9\textwidth]{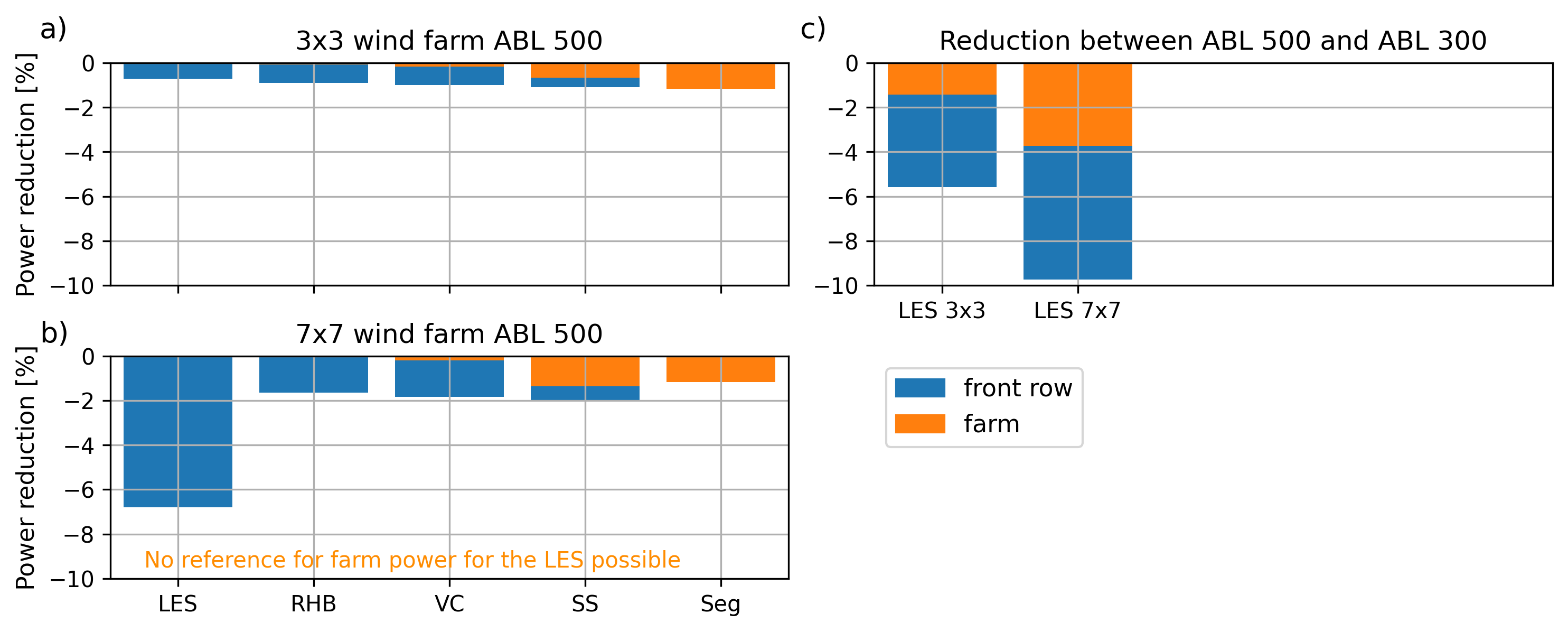}
\caption{Reduction of the average power production of the first row and the whole wind farm caused by modeling the induction effect, for farm 3$\times$3 a) and 7$\times$7 b). The references are the same simulations without modelling any induction effects. For the LES, the reference of the first row power is a simulation of the isolated first row. For the whole farm power no reference is possible for the LES case. c) The reduction of the average power production of the first row and the whole wind farm for the two different wind farms in the simulation setups with different boundary layer height. The reference case is the setup ABL 500.}
\label{fig:power}
\end{figure}

\subsection{Wind Farm power}\label{sec:wf_pow}
In the previous section, the comparison with the LES has shown the linear superposition of induction models to be far from sufficient to describe the velocity deficit caused by global blockage. The enhancement of the flow visualization may not be a sufficient reason justifying the method deployment in context such as \ac{AEP} assessment or wind farm optimization algorithms. The many degrees of freedom involved in the solution of these problems lead to iterative procedures that require fast calculation of the farm output at a given state. Any computational cost added to this task must be compensated by clear improvements in the accuracy of the prediction.
In the following section, we debate the benefits of considering an induction model in the assessment of wind farm production.

In Fig.\,\ref{fig:power}(a,b) we show the percentage of power reduction at the first row and to the whole farm, by the application of the induction models and the global wind speed reduction at any of the farm's turbine, according to Eq. \ref{eq:Seg}.
Because induction in the LES cannot be switched off, we define its reference as the results for power output of a simulation where a single row of the farm is considered. This practice is identical to the reference taken for the analytical models, as removing induction in this models is equivalent to removing the effects of the downwind turbines. Such a normalization for the LES is also beneficial to remove row effects that lead to higher power production as shown in \cite{MeyerForsting2017}. 
Unfortunately, it is not possible to define a similar reference for the whole farm in the LES.

The trend of velocity reduction observed in Fig. \ref{fig:ws_x_avg}, can also be observed in the first row power predicted by the LES, that decreases evidently as the farm size grows and the \ac{ABL} height decreases. From the 3$\times$3 to the 7$\times$7 the power reduction grows of 14 times in the LES at ABL 500, while in the models it merely double. A further evidence of a strongly non linear phenomena. The models also consistently underestimate the power reduction at first row for all the studied cases but the 3$\times$3\_500, in which the effect of flow blockage are minimal. This result shows that the inlet velocity deficit may be partially recovered, another sign of the importance of the pressure effects. 

The power production of the whole farm is less affected by the global blockage effect than the first row in all the simulations with the induction models. This result is also confirmed by the LES. The relative power reduction across the two \ac{ABL} simulated is much greater for the first row than for the whole farm, Fig. \ref{fig:power}c, indicating as in Fig. \ref{fig:ws_x_avg}  that the inflow deficit is recovered at the downwind turbines. We therefore urge caution with use of approaches based on reducing the velocity at all the turbines in the farm according to a correction measured upstream, e.g. Eq. \ref{eq:Seg}. Despite we could not prove it here, we expect the methods to be biased towards underestimating the whole farm power. 

To conclude the results' section, we finally discuss the impact induction models have on the whole farm power production. In our simulations, the use of neither the \ac{VC} model nor the RHB model significantly affect this prediction. To our understanding, this is connected to the speed up around the rotor described in these models, that despite being subtle it is sufficient to balance the negative effect of induction on power. 
To this extent, only the \emph{Self-Similar} model provides a prediction of the order of magnitude typically assumed for the global blockage effect. More research in this direction is still necessary but for the other two models, the significant computational cost we experienced in our simulations, especially with the \ac{VC}, and the marginal influence on power production are results that do not encourage their use in \ac{AEP} calculation. 

\section{A simple dimensional analysis of global blockage effect}\label{s:dimension}

According to our analysis, we find that wind farm size, as well as atmospheric boundary layer height, have a significant impact on the extent of velocity reduction in front of the wind farm, see Fig. \ref{fig:ws_x_avg}. This motivates us to perform a simple dimensional analysis to find out whether a universal behaviour of the velocity deficit can be identified. \\
Our LES simulations suggest that the wind speed reduction upstream the farm, here defined as $u_{x,loss} = U_{x,ref} - U_{x,LES}$, can be represented as a firt approach by the functional relation 
\begin{equation}\label{eq:funrel}
u_{x,loss} = f(x,U_{x,ref},L_{farm},H_{abl}). 
\end{equation}
In Eq. \ref{eq:funrel}, $x$ denotes the axial coordinate with $x=0$ at the farm first row, $U_{x,ref}$ the undisturbed velocity, $L_{farm}$ the length of the farm, and $H_{abl}$ the height of the \ac{ABL}. We decide to neglect the width of the farm, as having fixed the aspect ratio in the two farms simulated, $L_{farm}$ is sufficient to describe their geometry change.

Casting the problem in the context of the Buckingham-Vaschy $\pi$-\,theorem \cite{Vaschy1892,Buckingham1914}, 
there are five total variables, and two fundamental dimensions, length and time, therefore three $\pi$ parameters must be defined.
\begin{align*}
\centering
\pi_u = \frac{u_{x,loss}}{U_{x,ref}} \quad\quad \pi_x = \frac{x}{L_{farm}} \quad\quad \pi_h = \frac{H_{abl} - D}{L_{farm}}
\end{align*}

\begin{figure}[t]
\centering
\begin{tikzpicture}[spy using outlines={rectangle,black,magnification=2, connect spies}]
\node {\pgfimage[interpolate=true, width=0.5\textwidth]{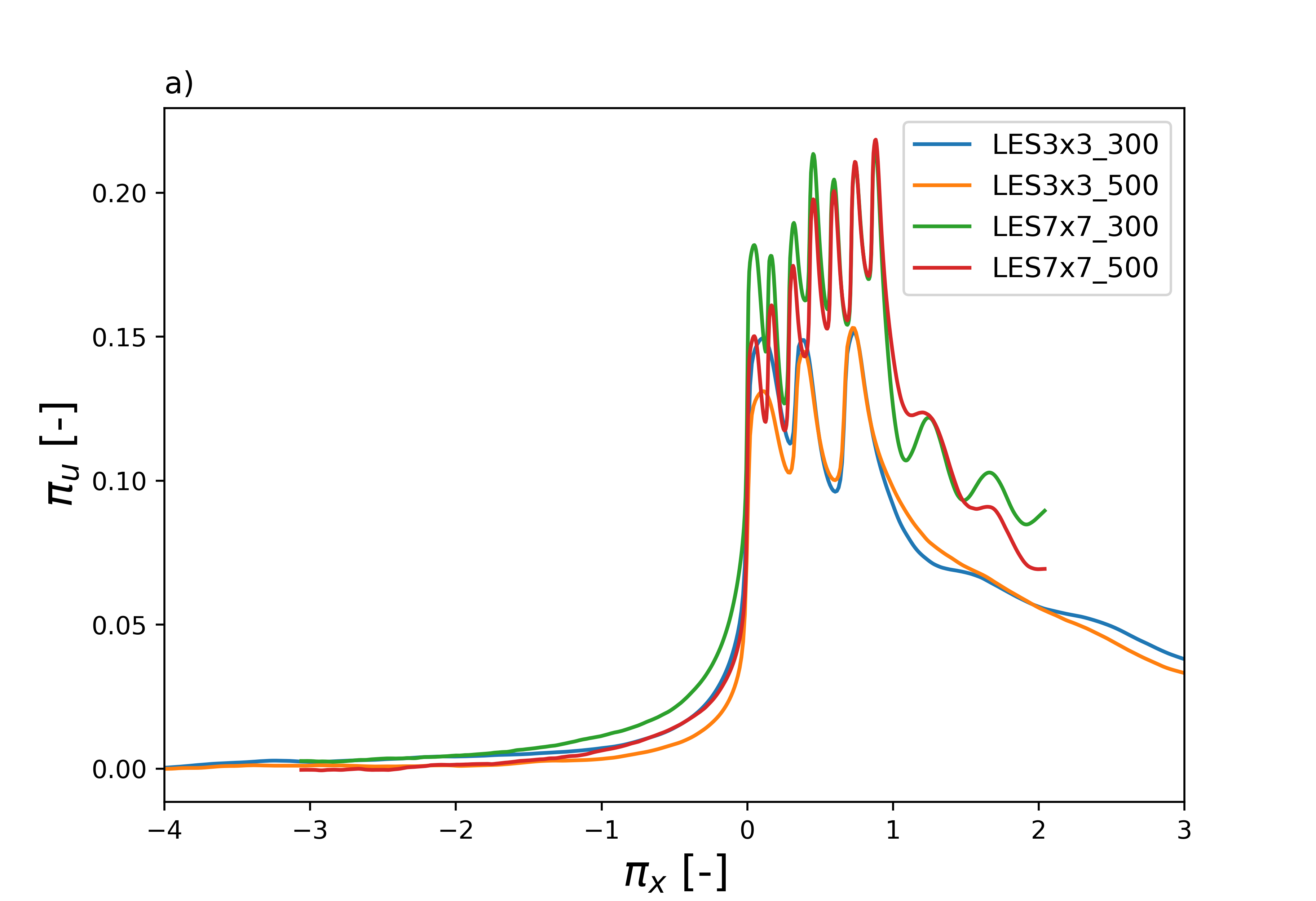}};
\spy[rectangle, black, width=3.2cm, height=2.5cm] on (-0.15,-1.4) in node [left] at (0.3,0.71);
\end{tikzpicture}%
\begin{tikzpicture}[spy using outlines={rectangle,black,magnification=2, connect spies}]
\node {\pgfimage[interpolate=true,width=0.5\textwidth]{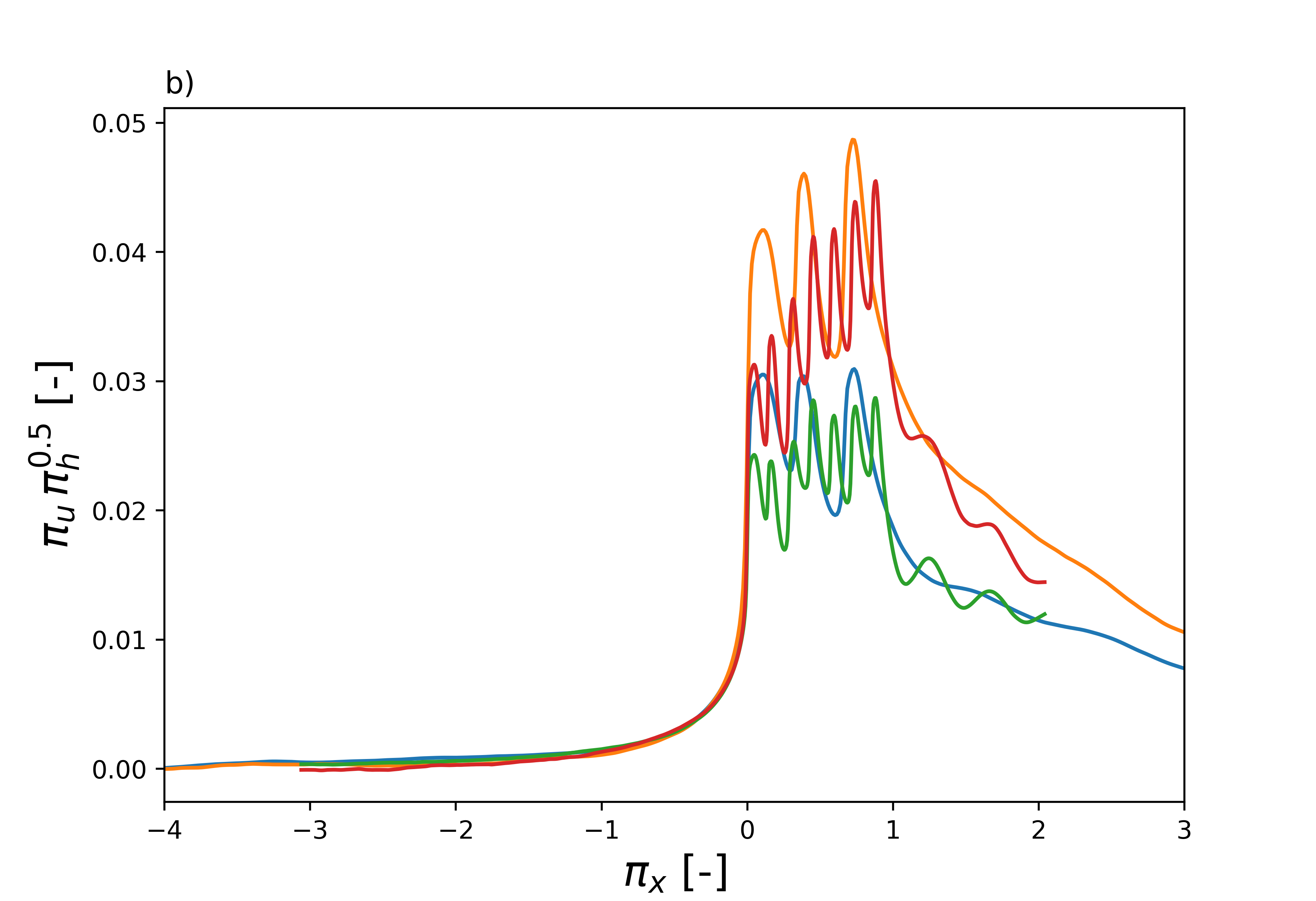}};\spy[rectangle, black, width=3.2cm, height=2.5cm] on (-0.15,-1.4) in node [left] at (0.3,0.71);

\end{tikzpicture}
\caption{Profiles of normalized loss velocity $\pi_u = u_{x,loss}/U_{ref}$ at hub height in spanwise average, along the full LES domain $x$, here normalized as $\pi_x = x/L_{farm}$. No rescaling for $\pi_u$ a), rescaling with $(\frac{H_{abl} - D}{L_{farm}})^{-0.5} b)$}
\label{fig:powerlaw}
\end{figure}
The choice of the first two groups appears as the most convenient, the one for the last group deserves a further comment. From our understanding, the wind can adapt to the presence of the farm by deflecting upward, downward, and to the sides of the area occupied by the rotors. As any deflection to the sides comes virtually with no consequence, the deflection up and down can cause extra shear as the flow is bounded by the ground and the capping inversion. 
The parameter $\pi_h$ is therefore representing the height of the channels upwards and downwards the wind farm rotors, whose cumulative height is $H_{abl} - D$. In terms of the $\pi$- theorem, we have incorporated the variable $D$ (the rotor diameter) only here to avoid an additional $\pi$- term. This choice, taken for sake of simplicity, is also justified by the fact that $D$ is constant across the LES, therefore we could have not investigated the dependence on this parameter.

According to the new non-dimensional $\pi$-\,terms, Eq.\,(\ref{eq:funrel}) can be rewritten as,

\begin{equation}\label{eq:adi}
    \pi_u = F\left( \pi_x,\,\pi_h\right)
\end{equation}

Mathematically, we now propose that $F$ factorises to $F\left( \pi_x,\,\pi_h\right) = G\left( \pi_x\right) g(\pi_h)$. As the LES results provide us the trend of $\pi_u(\pi_x)$, we search for the functional relation $g(\pi_h)$, so that
\begin{equation}
    \frac{\pi_u}{g(\pi_h)} = G\left( \pi_x\right) 
\end{equation}
with $G(\pi_x)$ being the universal curve describing the velocity deficit. 
As demonstrated in the right figure of (\ref{fig:powerlaw}), choosing 
\begin{equation}
    g(\pi_h) = \pi_h^{-0.5}
\end{equation}
gains the four curves predicted by the LES, Fig. \ref{fig:powerlaw} (left), to collapse to one in the inflow region.

It must be underlined that this is only a first and early attempt at characterising the interactions of wind farm and ABL we believe to originate the GBE. It is interesting that this simple application of the $\pi$-\,theorem provided a rather promising result. Lacking a more substantial set of observations to test the model on, we avoid trying to interpret to much out of it. But we  believe the whole modelling of global blockage would benefit from more work in this direction.

\section{Conclusion}

By means of comparison with LES, the analytic induction models are shown to correctly describe the induction generated by a single wind turbine in isolation and to improve the flow representation within the wind farm, especially if the model accounts for lateral speed-ups.
Under the conditions simulated, these results do not carry over when these models are linearly superimposed to describe the wind farm inflow deceleration due to global blockage. We observe a substantial underestimation in the model predictions that increases as the wind farm dimension grows and the ABL height decreases. These results show the poor sensitivity, or complete lack thereof, to these parameters to generate large inaccuracies in the models predictions.

We deduce that the deceleration of the inflow is mainly determined by the interactions and the physics of the ABL rather than by the turbines' induction. Compelling evidence at support of this thesis are also found in the production data. We observe the inflow deficit to be partially recovered at first row and especially at the row downstream in the farm. Suggesting the global blockage effect to be not correlated to the turbines induction but rather to exogenous pressure perturbations, likely due to the displacement of air masses with different density aloft \cite{Allaerts2017}, capable of redistributing the available energy.
We conclude that methods to predict the GBE at wind farm based on reducing the whole farm speed velocity according to the measured inflow deficit, e.g. as in Eq.\,(\ref{eq:Seg}), are not advisable. We also believe the linear superposition of induction models to be inadequate to the modeling of GBE in its present form where no dynamics of the ABL are considered.

Eventually, we perform a dimensional analysis of the inlet velocity deficit predicted by the LES aimed at simplifying the description of the interactions between wind farm and atmospheric boundary layer. Despite being very early, the results are promising and they could give birth to a new approach for the modeling of the GBE.

Future work is necessary to more broadly characterize the extent of the GBE at varying conditions of the ABL, and the wind farm size and shape. This will gain a better testing ground for the presented dimensional model, and nourish further development of the approach.
In parallel, we foresee the modeling of the lower atmosphere's vertical structure, such the one presented in \cite{Allaerts2019}, as a possible step forward of the current status.

\begin{acronym}
\acro{RANS}{Reynolds-Averaged Navier-Stokes}
\acro{LES}{Large-Eddy Simulation}
\acro{AD}{actuator disk}
\acro{WT}{wind turbine}
\acro{VC}{Vortex Cylinder}
\acro{RHB}{Rankine Half-Body}
\acro{SS}{Self-Similar}
\acro{ABL}{atmospheric boundary layer}
\acro{GBE}{global blockage effect}
\acro{AEP}{Annual Energy Production}
\acro{EM}{engineering models}
\acro{SCADA}{Supervisory control and data acquisition}
\end{acronym}

\ack{The work presented in this paper was developed in the framework of the X-Wakes project (FKZ 03EE3008D,A) which is funded by the German Federal Ministry for Economic Affairs and Energy (Bundesministerium für Wirtschaft und Energie - BMWi) due to a decision of the German Bundestag. The simulations were partly performed at the HPC Cluster EDDY, located at the University of Oldenburg (Germany) and funded by BMWi (FKZ 0324005). Additional computer resources have been
provided by the North German Supercomputing Alliance (HLRN).
}
\section*{References}
\bibliography{references}
\end{document}